\documentclass[english,aps,tightenlines,showpacs, showkeys, notitlepage]{revtex4-2}
\usepackage[T1]{fontenc}
\usepackage[latin9]{inputenc}
\usepackage{color}
\usepackage{babel}
\usepackage{amstext}
\usepackage{esint}
\usepackage[unicode=true,pdfusetitle,
 bookmarks=true,bookmarksnumbered=false,bookmarksopen=false,
 breaklinks=false,pdfborder={0 0 0},pdfborderstyle={},backref=false,colorlinks=true]
 {hyperref}
\begin{document}
\title{Born-Infeld-$f\left(R\right)$ gravity with de Sitter solutions}
\author{Salih Kibaro\u{g}lu}
\email{salihkibaroglu@maltepe.edu.tr}

\date{\today}
\begin{abstract}
In this study, we consider the Born--Infeld-$f(R)$ gravity in which
the $f\left(R\right)$ term enters directly into the square root in
the Palatini formulation. We shortly analyzed this model for an explicit
$f\left(R\right)$ function which includes positive and negative powers
of the curvature scalar. We also show that both the ordinary Born--Infeld-$f(R)$
and this modification reduce to the same gravitational action form
under the conformal approach. Then we consider the existence of a
maximally symmetric vacuum solution for the gravitational field equations
and find the de Sitter solution for this modified model.
\end{abstract}
\affiliation{Maltepe University, Faculty of Engineering and Natural Sciences, 34857,
Istanbul, Turkey}
\affiliation{Institute of Space Sciences (CSIC-IEEC) C. Can Magrans s/n, 08193
Cerdanyola (Barcelona) Spain}
\maketitle

\section{Introduction}

Einstein's theory of general relativity (GR) is the most successful
theory that explains almost all the gravitational mechanisms for around
a century. The validity of this theory has been supported by lots
of observation and experiments on a wide range of scales where it
has been directly tested. Despite its observational success, there
are some reasons to believe that GR is unable to explain some gravitational
phenomena that need to be clarified. In particular, GR does not give
consistent results at the Planck scale where quantum effects play
an important role. In more general words, at extreme conditions such
as ultra-high or low energies, GR is expected to be modified.

In this sense, Born-Infeld (BI) like gravitational theories provide
a useful framework to modify GR. The original theory proposed by Born
and Infeld have a non-linear action for classical electrodynamics
\citep{born1934foundations}. In analogy with this study, Deser and
Gibbons established a gravitational model by using Born-Infeld like
determinantal structure using the Ricci tensor instead of the electromagnetic
field tensor \citep{deser1998born}. But this attempt remained unsuccessful
because of the presence of ghost fields. To resolve this problem,
Vollick succeeded to construct a theory without ghosts by using the
Palatini formulation which the connection and the metric tensor are
considered independent fields \citep{vollick2004palatini,vollick2005born}.
In 2010, Bañados and Ferreira made this theory in more standard form,
especially in the definition of the matter term \citep{Banados2010eddington}.
After these works, lots of applications and modifications of BI gravity
appeared in the literature (for a detailed review see \citep{jimenez2018born}). 

One of the important extension is BI-$f\left(R\right)$ \citep{makarenko2014born}.
In this model, the BI gravitational action is combined with the well-known
$f\left(R\right)$ theories in the Palatini formalism. This model
attracted a lot of attention and it is used wide variety of fields
especially in cosmology \citep{odintsov2014born,Makarenko:2014nca,makarenko2014unification,Makarenko:2014cca,Elizalde:2016vsd,chen2016modified,banik2018dynamical}.
On the other hand there is different generalization of BI-$f\left(R\right)$
model which $f\left(R\right)$ function directly enters in the determinantal
structure \citep{makarenko2014born,chen2016modified}. This model
has been studied less than the other and it is capable to provide
a different framework \citep{chen2016modified}. 

From this background, the main purpose of this study is to analyze
BI inspired model of gravity coupled to a $f\left(R\right)$ term
which is presence in the determinantal structure. In this context,
we first give a detailed review of the current situation for this
model. Besides we analyze the de Sitter solution which is very important
when we consider inflationary period of our universe \citep{guth1981inflationary,linde1982new,albrecht1982cosmology,linde1983chaotic,linde1990particle,linde1991axions,linde1994hybrid,kallosh2002gauged}
for this model. 

In Section II, we give a brief review of the Born-Infeld-$f\left(R\right)$
gravity. In Section III, a different approximation of the BI-$f\left(R\right)$
gravity which the $f\left(R\right)$ term directly enter into the
determinant is given. We also examine three cases for this extension.
In the first case, we use an exact $f\left(R\right)$ function which
has positive and negative powers of the Ricci scalar. In the second
case, we analyze the conformal approximation in the metric tensor.
In the last one, we derive the de Sitter solution for this model.
In the last section, we conclude the paper with some discussion.

\section{Born-Infeld-$f\left(R\right)$ theory}

Let us briefly review the standard BI-$f\left(R\right)$ theory \citep{makarenko2014born}
(see also \citep{Makarenko:2014nca,odintsov2014born,makarenko2014unification}).
In this theory, the original EiBI gravity theory is combined with
an additional $f\left(R\right)$ that depends on the Ricci scalar
$R=g^{\mu\nu}R_{\mu\nu}\left(\Gamma\right)$. To avoid any ghost instabilities,
the theory is formulated within the Palatini formalism (for more detail
see \citep{Olmo:2011uz,olmo2012open}), in which the metric $g_{\mu\nu}$
and the connection $\Gamma_{\beta\gamma}^{\alpha}$ are treated as
independent variables. The action for this theory is given by

\begin{eqnarray}
S & = & \frac{2}{\kappa}\int d^{4}x\left[\sqrt{-|g_{\mu\nu}+\epsilon R_{\mu\nu}|}-\lambda\sqrt{-g}\right]+\int d^{4}x\left[\sqrt{-g}f\left(R\right)\right]+S_{m},\label{eq: action_f(R)}
\end{eqnarray}
where the first term represents the standard BI gravitational Lagrangian
and the second term is an additional function of the Ricci scalar.
$S_{m}$ is the matter action which depends on generically $\psi$
field and the metric tensor $g_{\mu\nu}$.

\begin{equation}
R_{\mu\beta\nu}^{\alpha}=\partial_{\beta}\Gamma_{\nu\mu}^{\alpha}-\partial_{\nu}\Gamma_{\mu\beta}^{\alpha}+\Gamma_{\beta\lambda}^{\alpha}\Gamma_{\nu\mu}^{\lambda}-\Gamma_{\nu\lambda}^{\alpha}\Gamma_{\mu\beta}^{\lambda},
\end{equation}
is the Riemann tensor of the connection $\Gamma_{\mu\nu}^{\lambda}$.
In addition, the connection is also assumed to be torsionless. The
parameter $\kappa$ is a constant with inverse dimensions to that
of a cosmological constant and $\lambda$ is a dimensionless constant.
Note that $g$ is the determinant of the metric. Throughout this paper,
we will use Planck units $8\pi G=1$ and set the speed of light to
$c=1$. 

The variation of this action with respect to the metric tensor leads
to a modified metric field equations for the standard BI gravitational
model,
\begin{equation}
\frac{\sqrt{-q}}{\sqrt{-g}}\left(q^{-1}\right)^{\mu\nu}-\lambda g^{\mu\nu}+\frac{\kappa}{2}g^{\mu\nu}f\left(R\right)-\kappa f_{R}R^{\mu\nu}=-\kappa T^{\mu\nu},\label{eq: eom_g-1}
\end{equation}
where $f_{R}$ is the derivative of $f\left(R\right)$ with respect
to the Ricci scalar, and $T^{\mu\nu}$ is the standard energy-momentum
tensor. We have used the notation,
\begin{equation}
q_{\mu\nu}=g_{\mu\nu}+\kappa R_{\mu\nu}.\label{eq: q_f(R)-1}
\end{equation}
We denoted the inverse of $q_{\mu\nu}$ by $\left(q^{-1}\right)^{\mu\nu}$
and $q$ represents the determinant of $q_{\mu\nu}$. Similarly, the
corresponding equation which follows by variation over the connection
$\Gamma$ has the form,

\begin{equation}
\nabla_{\lambda}\left(\sqrt{-q}q^{\mu\nu}+\sqrt{-g}f_{R}g^{\mu\nu}\right)=0,\label{eq: eom_conn-1}
\end{equation}
where the covariant derivative is taken with respect to the connection
which is defined for a scalar field $\phi$ as
\begin{equation}
\nabla_{\mu}\phi=\partial_{\mu}\phi-\Gamma_{\mu\alpha}^{\alpha}\phi.
\end{equation}
If we assume that $q_{\mu\nu}$ is conformally proportional to the
metric tensor as 
\begin{equation}
q_{\mu\nu}=k\left(t\right)g_{\mu\nu},\label{eq: conf_rel}
\end{equation}
 then Eq.(\ref{eq: eom_conn-1}) becomes,
\begin{equation}
\nabla_{\nu}\left[\sqrt{-u}\left(u^{-1}\right)^{\mu\nu}\right]=0.\label{eq: eom_u}
\end{equation}
In this case, we have an auxiliary metric defined as $u_{\mu\nu}=\left[k\left(t\right)+f_{R}\right]g_{\mu\nu}$
and $\left(u^{-1}\right)^{\mu\nu}$ represents the inverse representation
of $u_{\mu\nu}$. Eq.(\ref{eq: eom_u}) tells us that the connection
can be defined by this auxiliary metric as
\begin{equation}
\Gamma_{\mu\nu}^{\rho}=\frac{1}{2}\left(u^{-1}\right)^{\rho\sigma}\left(u_{\sigma\nu,\mu}+u_{\mu\sigma,\nu}-u_{\mu\nu,\sigma}\right).
\end{equation}
By considering the conformal relation between $q_{\mu\nu}$ and $g_{\mu\nu}$
in Eq.(\ref{eq: conf_rel}), we can easily say that the Ricci tensor
must also be proportional to the metric tensor. Thus one can write
the relationship between the Ricci tensor and $g_{\mu\nu}$, as
\begin{equation}
R_{\mu\nu}=\frac{1}{\kappa}\left[k\left(t\right)-1\right]g_{\mu\nu}.\label{eq: Ricci_g}
\end{equation}

For a cosmological scenario, let us consider a homogeneous and isotropic
the Friedman--Lemaitre--Robertson--Walker (FLRW) universe with
metric,
\begin{equation}
ds^{2}=-dt^{2}+a^{2}\left(t\right)\left(dx^{2}+dy^{2}+dz^{2}\right),\label{eq: metric_FRW}
\end{equation}
where $t$ is the cosmic time and $a\left(t\right)$ is the scale
factor. Now, we can define the auxiliary metric as
\begin{equation}
u_{\mu\nu}=u\left(t\right)\text{diag}\left(-1,a^{2}\left(t\right),a^{2}\left(t\right),a^{2}\left(t\right)\right),\label{eq: metric_U}
\end{equation}
where $u\left(t\right)=k\left(t\right)+f_{R}$. According to Eq.(\ref{eq: Ricci_g}),
we can define the Ricci tensor as $R_{\mu\nu}=r\left(t\right)g_{\mu\nu}$.
From this background, one can find the following expressions
\begin{equation}
r\left(t\right)=3\left(H+\frac{\dot{u}}{2u}\right)^{2},\label{eq: r_1}
\end{equation}
\begin{equation}
2\dot{H}=H\frac{\dot{u}}{u}+\frac{3}{2}\left(\frac{\dot{u}}{u}\right)^{2}-\frac{\ddot{u}}{u},
\end{equation}
where the upper dot denotes the time derivative and $H=\frac{\dot{a}}{a}$
is the Hubble parameter. Using these equations, one can derive the
following relation
\begin{equation}
u\left(t\right)=cr\left(t\right)\label{eq: u_r_relation}
\end{equation}
where $c$ is an integration constant. Then combining Eq.(\ref{eq: r_1})
and Eq.(\ref{eq: u_r_relation}), one can get
\begin{equation}
H=\pm\sqrt{\frac{u}{3c}}-\frac{\dot{u}}{2u},\label{eq: Hubble_parameter}
\end{equation}
and Eq.(\ref{eq: u_r_relation}) lead to find the $f\left(R\right)$
function as
\begin{equation}
f\left(R\right)=\Lambda-R+\left(\frac{c-\kappa}{8}\right)R^{2},
\end{equation}
where the effective cosmological constant is defined by $\Lambda=\frac{\lambda-1}{\kappa}$.
If we substitute this function into the action Eq.(\ref{eq: action_f(R)})
without matter field, we obtain the following action by setting the
constants $\kappa c=16$.
\begin{equation}
S=\frac{2}{\kappa}\int d^{4}x\sqrt{-g}R^{2}.\label{eq: action_2}
\end{equation}

\section{$F\left(R\right)$ term in the determinant}

In this section, we examine a modified BI gravitational theory with
a function of the Ricci scalar being added to the determinantal action.
This model has been partly analyzed in \citep{makarenko2014born,chen2016modified}
(for early development similar to this model in pure metric formalism
\citep{comelli2005determinant} and in the teleparallel framework\citep{Fiorini:2013kba}).
We use the construction method for the BI-like action given in \citep{chen2016modified}
to realize our purpose. To this aim, we use the following action 

\begin{eqnarray}
S & = & \frac{2}{\kappa}\int d^{4}x\left[\sqrt{-|g_{\mu\nu}+\kappa F_{\mu\nu}|}-\lambda\sqrt{-|g_{\mu\nu}|}\right]+S_{m},\label{eq: action_f(r)_2}
\end{eqnarray}
here $F_{\mu\nu}$ is defined as follows,

\begin{equation}
F_{\mu\nu}=\alpha R_{\mu\nu}\left(\Gamma\right)+\beta g_{\mu\nu}F\left(R\right),
\end{equation}
where $\alpha$ and $\beta$ are dimensionless constants. To find
the field equations, we first define a new object as
\begin{equation}
p_{\mu\nu}=g_{\mu\nu}+\kappa\left[\alpha R_{\mu\nu}+\beta g_{\mu\nu}F\left(R\right)\right],\label{eq: p_f(R)}
\end{equation}
 here the inverse of $p_{\mu\nu}$ is denoted $\left(p^{-1}\right)^{\mu\nu}$
and these objects satisfy $\left(p^{-1}\right)^{\mu\rho}p_{\rho\nu}=\gamma_{\nu}^{\mu}$.
By using Eq.(\ref{eq: p_f(R)}), the action Eq.(\ref{eq: action_f(r)_2})
reads
\begin{eqnarray}
S & = & \frac{2}{\kappa}\int d^{4}x\left[\sqrt{-p}-\lambda\sqrt{-g}+\frac{\kappa}{2}\mathcal{L}_{m}\right],
\end{eqnarray}
where $p=det\left(p_{\mu\nu}\right)$. To obtain equations of motion,
we look the variation with respect to metric $g_{\mu\nu}$ and the
connection $\Gamma_{\beta\mu}^{\alpha}$, respectively,

\begin{equation}
\sqrt{-p}\left(p^{-1}\right)^{\mu\nu}\left(1+\kappa\beta F\left(R\right)\right)-\kappa\beta\sqrt{-p}\left(p^{-1}\right)^{\sigma\rho}g_{\sigma\rho}F_{R}g^{\mu\alpha}g^{\nu\beta}R_{\alpha\beta}-\lambda\sqrt{-g}g^{\mu\nu}=-\kappa\sqrt{-g}T^{\mu\nu},\label{eq: eom_g}
\end{equation}

\begin{equation}
\nabla_{\lambda}\left[\sqrt{-p}\left(\alpha\left(p^{-1}\right)^{\mu\nu}+\beta\left(p^{-1}\right)^{\sigma\rho}g_{\sigma\rho}g^{\mu\nu}F_{R}\right)\right]=0.\label{eq: eom_conn}
\end{equation}

According to the literature, it is well known that the Palatini $F\left(R\right)$
theories can be solved in terms of an auxiliary metric. From this
idea and taking account of Eq.(\ref{eq: eom_conn}), we define an
auxiliary metric $u_{\mu\nu}$ as follows 

\begin{equation}
\sqrt{-p}\left(\alpha\left(p^{-1}\right)^{\mu\nu}+\beta\left(p^{-1}\right)^{\sigma\rho}g_{\sigma\rho}g^{\mu\nu}f_{R}\right)=\sqrt{-u}\left(u^{-1}\right)^{\mu\nu},\label{eq: def_u}
\end{equation}
where $u=det\left(u_{\mu\nu}\right)$ and $\left(u^{-1}\right)^{\mu\nu}$
is the inverse of $u_{\mu\nu}$, that is $\left(u^{-1}\right)^{\mu\rho}u_{\rho\nu}=\delta_{\nu}^{\mu}$.
By considering the determinant of Eq.(\ref{eq: def_u}), we find,
\begin{equation}
p^{2}det\left(\alpha\left(p^{-1}\right)^{\mu\nu}+\beta\left(p^{-1}\right)^{\sigma\rho}g_{\sigma\rho}g^{\mu\nu}F_{R}\right)=u,
\end{equation}
and this definition leads to
\begin{eqnarray}
\left(u^{-1}\right)^{\mu\nu} & = & \frac{\left(\alpha p^{\mu\nu}+\beta p^{\sigma\rho}g_{\sigma\rho}g^{\mu\nu}F_{R}\right)}{\sqrt{det\left(\alpha\delta_{\xi}^{\zeta}+\beta p^{\sigma\rho}g_{\sigma\rho}p_{\xi}^{\zeta}F_{R}\right)}}.\label{eq: u_munu}
\end{eqnarray}
Now Eq.(\ref{eq: eom_conn}) can be rewritten
\begin{equation}
\nabla_{\lambda}\left[\sqrt{-u}\left(u^{-1}\right)^{\mu\nu}\right]=0,\label{eq: eom_u-1}
\end{equation}
which can be solved with respect to the connection $\Gamma_{\mu\nu}^{\rho}$
as in the Einstein gravity as follows,
\begin{equation}
\Gamma_{\mu\nu}^{\rho}=\frac{1}{2}\left(u^{-1}\right)^{\rho\sigma}\left(u_{\sigma\nu,\mu}+u_{\mu\sigma,\nu}-u_{\mu\nu,\sigma}\right).\label{eq: gamma_u}
\end{equation}

\subsection{An explicit $F\left(R\right)$ function}

In the low energy limit ($\kappa\rightarrow0$), the gravitational
action takes the following form
\begin{eqnarray}
S & = & \int d^{4}x\sqrt{-g}\left\{ \left(\alpha R+4\beta F\left(R\right)\right)+\kappa\left(\frac{\alpha^{2}}{4}R^{2}+\alpha\beta RF\left(R\right)+2\beta^{2}F\left(R\right)^{2}\right)-\frac{\kappa\alpha^{2}}{2}R_{\mu\nu}R^{\mu\nu}-2\Lambda+\mathcal{O}^{3}\left(R\right)\right\} +S_{m}.
\end{eqnarray}
Let us analyze an explicit $F\left(R\right)$ function which include
both positive and negative powers. This kind of theories have been
use in various kind of application in the context of gravitation \citep{nojiri2003modified,chiba20031,dolgov2003can,fay2007f,nojiri2007introduction,nojiri2011unified}.
We assume that the function has the following form,
\begin{equation}
F\left(R\right)=\gamma R+\frac{\epsilon}{R}.
\end{equation}
Firstly, we choose $\gamma=\epsilon=1$ and $\alpha+4\beta=1$ then
the action becomes

\begin{equation}
S=\int d^{4}x\sqrt{-g}\left\{ R+\frac{1-\alpha}{R}+\frac{\kappa\left(1-\alpha\right)^{2}}{8R^{2}}+\frac{\kappa\left(1+\alpha^{2}\right)}{8}R^{2}+\frac{\kappa\left(1-\alpha\right)}{4}-\frac{\kappa\alpha^{2}}{2}R_{\mu\nu}R^{\mu\nu}-2\Lambda\right\} +S_{m},\label{eq: action_FR_1}
\end{equation}
here we obtain an action with Einstein's theory for low curvature
limit together with both $R^{2}$ term and inverse power of $R$.
Secondly, if we choose $\gamma=1$ and $\epsilon=0$ we reach $F\left(R\right)=R$
then the action reduces to
\begin{equation}
S=\int d^{4}x\sqrt{-g}\left\{ \left(\alpha+4\beta\right)R+\kappa\left(\frac{\alpha^{2}}{4}+\alpha\beta+2\beta^{2}\right)R^{2}-\frac{\kappa\alpha^{2}}{2}\left(R_{\mu\nu}R^{\mu\nu}\right)-2\Lambda\right\} +S_{m}.\label{eq: action_FR_2}
\end{equation}
In this case, to recover General Relativity, the parameters should
satisfy again $\alpha+4\beta=1$. Moreover, if we set $\alpha=0$
and $\beta=\frac{1}{4}$, we derive $R^{2}$ theory of gravity as
\begin{equation}
S=\int d^{4}x\sqrt{-g}\left(R+\frac{\kappa}{8}R^{2}-2\Lambda\right)+S_{m}.
\end{equation}
This model has already been studied in \citep{chen2016modified}.
Thirdly, choosing $\gamma=0$ and $\epsilon=1$ we obtain $F\left(R\right)=1/R$
and assuming $\alpha=1$, $\beta=1/4$ then the action takes the following
form

\begin{equation}
S=\int d^{4}x\sqrt{-g}\left\{ R+\frac{1}{R}+\frac{\kappa}{8R^{2}}+\frac{\kappa}{4}R^{2}+\frac{\kappa}{4}-\frac{\kappa}{2}\left(R_{\mu\nu}R^{\mu\nu}\right)-2\Lambda\right\} +S_{m}.\label{eq: action_FR_3}
\end{equation}
For the last case, the selection $\alpha=1$ and $\beta=0$ goes to
standard formulation of the BI gravity. 

\subsection{Conformal case}

The conformal assumption of this model has been already examined in
\citep{makarenko2014born}. Here, we wish to analyze this case for
our notation. Let us start to give the main definition for this case
if we assume the conformal relationship between $p_{\mu\nu}$ and
$g_{\mu\nu}$ as
\begin{equation}
p_{\mu\nu}=f\left(t\right)g_{\mu\nu},\label{eq: p_conf}
\end{equation}
then Eq.(\ref{eq: def_u}) takes the following form,
\begin{equation}
f\left(t\right)\left(\alpha+4\beta F_{R}\right)\sqrt{-g}g^{\mu\nu}=\sqrt{-u}\left(u^{-1}\right)^{\mu\nu}.
\end{equation}
From this definition, we can easily write
\begin{equation}
u_{\mu\nu}=\left[f\left(t\right)\left(\alpha+4\beta F_{R}\right)\right]g_{\mu\nu}.\label{eq: func_u}
\end{equation}
In this case, the conditions are satisfied given in Eq.(\ref{eq: eom_u-1})
and Eq.(\ref{eq: gamma_u}). The conformal case (\ref{eq: p_conf})
leads to write a condition in which the Ricci tensor is proportional
to the metric tensor $g_{\mu\nu}$ as,
\begin{equation}
R_{\mu\nu}=r\left(t\right)g_{\mu\nu},\label{eq: Ricci_g_relation}
\end{equation}
where $r\left(t\right)$ can be found by using Eq.(\ref{eq: p_f(R)})
and Eq.(\ref{eq: p_conf}) as
\begin{equation}
r\left(t\right)=\frac{1}{\kappa\alpha}\left[f\left(t\right)-1-\kappa\beta F\left(R\right)\right].\label{eq: r_2}
\end{equation}
Furthermore, taking trace of Eq.(\ref{eq: Ricci_g_relation}), one
can also write $r\left(t\right)=\frac{R}{4}$ and Eq.(\ref{eq: Ricci_g_relation})
reduces to
\begin{equation}
R_{\mu\nu}=\frac{R}{4}g_{\mu\nu}.
\end{equation}

Now if we suppose that for the spatially-flat FLRW universe with metric
Eq.(\ref{eq: metric_FRW}) and the definition of $u_{\mu\nu}$ in
Eq.(\ref{eq: metric_U}), we find same relationship as given in Eq.(\ref{eq: u_r_relation})
and this model also satisfies Eq.(\ref{eq: Hubble_parameter}). By
the help of Eq.(\ref{eq: u_r_relation}) and Eq.(\ref{eq: func_u}),
we find the $F\left(R\right)$ function as the following form

\begin{equation}
F\left(R\right)=-\frac{1}{\kappa\beta}-\frac{\alpha}{4\beta}R\pm\frac{\sqrt{cR^{2}+\frac{1}{c_{1}^{2}\kappa^{2}\beta^{2}}}}{4\kappa\beta},\label{eq: F(R)-1}
\end{equation}
where $c_{1}$ is an integration constant. So substituting Eq.(\ref{eq: F(R)-1})
into (\ref{eq: action_f(R)}) together with setting $c=16$ and $\lambda=\frac{1}{16c_{1}^{2}\kappa^{2}\beta^{2}}$,
the action takes the same form as given before in Eq.(\ref{eq: action_2}).

\subsection{The de Sitter solution}

In this part, our investigation into the existence of the de Sitter
vacuum solution. This solution provides a very useful theoretical
background for early universe exponential expansion. This model is
described by a constant Hubble parameter and from this point, the
Riemann tensor can be written
\begin{equation}
R_{\alpha\beta\rho\sigma}=\frac{1}{12}R\left(g_{\alpha\rho}g_{\beta\sigma}-g_{\alpha\sigma}g_{\beta\rho}\right).
\end{equation}
Taking account of the Bianchi identities, we can write $R=R_{0}$
as a covariantly constant curvature. So the Ricci tensor satisfies,

\begin{equation}
R_{\mu\nu}=\frac{R_{0}}{4}g_{\mu\nu}.\label{eq: cons_curvature}
\end{equation}
Then Eq.(\ref{eq: p_f(R)}) becomes,
\begin{equation}
p_{\mu\nu}=\left\{ 1+\kappa\left[\frac{\alpha}{4}R_{0}+\beta F\left(R_{0}\right)\right]\right\} g_{\mu\nu},
\end{equation}

\begin{equation}
\left(p^{-1}\right)^{\mu\nu}=\left\{ 1+\kappa\left[\frac{\alpha}{4}R_{0}+\beta F\left(R_{0}\right)\right]\right\} ^{-1}g^{\mu\nu}.
\end{equation}
If we apply these results to (\ref{eq: def_u}), we get
\begin{eqnarray}
\sqrt{-u}\left(u^{-1}\right)^{\mu\nu} & = & \sqrt{-g}\left[\left(\alpha+4\beta F_{R_{0}}\right)\left(1+\frac{\kappa\alpha}{4}R_{0}+\kappa\beta F\left(R_{0}\right)\right)\right]g^{\mu\nu},
\end{eqnarray}
where $F_{R_{0}}$ is the derivative with respect to $R_{0}$. Now
$\left(u^{-1}\right)^{\mu\nu}$ takes the following form

\begin{equation}
\left(u^{-1}\right)^{\mu\nu}=\left(\alpha+4\beta F_{R_{0}}\right)^{-1}\left(1+\frac{\kappa\alpha}{4}R_{0}+\kappa\beta F\left(R_{0}\right)\right)^{-1}g^{\mu\nu},\label{eq: u_inverse}
\end{equation}
which tells, by using (\ref{eq: gamma_u}), we find that the connection
$\Gamma_{\mu\nu}^{\rho}$ is nothing but that of the Einstein gravity
\begin{equation}
\Gamma_{\mu\nu}^{\rho}=\frac{1}{2}g^{\rho\sigma}\left(g_{\sigma\nu,\mu}+g_{\mu\sigma,\nu}-g_{\mu\nu,\sigma}\right).\label{eq: gamma_u-1}
\end{equation}
Then by using Eq.(\ref{eq: u_inverse}), Eq.(\ref{eq: eom_g}) becomes

\begin{equation}
\left[1+\frac{\kappa\alpha}{4}R_{0}+\kappa\beta F\left(R_{0}\right)\right]\left[1+\kappa\beta F\left(R_{0}\right)-\kappa\beta F_{R_{0}}R_{0}\right]-\lambda=0,\label{eq: eom_g_2-1}
\end{equation}
where $T_{\mu\nu}=0$ because we are considering the vacuum (anti-)de
Sitter space-time. This is an algebraic equation with respect to $R_{0}$.
If the solution is positive, $R_{0}>0$, the solutions of Eq. (\ref{eq: cons_curvature})
are the de Sitter space-time, the de Sitter-Schwarzschild space-time,
and the de Sitter-Kerr space-time and if $R_{0}<0$, they are the
anti-de Sitter space-time, the anti-de Sitter-Schwarzschild space-time,
and the anti-de Sitter-Kerr space-time.

If we solve (\ref{eq: eom_g_2-1}) with respect to $F\left(R_{0}\right)$
we get
\begin{equation}
F\left(R_{0}\right)=-\frac{1}{\kappa\beta}-\frac{\alpha R_{0}}{4\beta}\pm\frac{\sqrt{c_{1}R_{0}^{2}+16\lambda}}{4\kappa\beta},\label{eq: F(R_0)}
\end{equation}
where $c_{1}$ is the integration constant. Now if we rearrange the
action (\ref{eq: action_f(r)_2}), we get

\begin{eqnarray}
S & = & \frac{2}{\kappa}\int d^{4}x\sqrt{-|g_{\mu\nu}|}\left\{ \left[1+\frac{\kappa\alpha}{4}R_{0}+\kappa\beta F\left(R_{0}\right)\right]^{2}-\lambda\right\} .\label{eq: action_desitter}
\end{eqnarray}
Inserting Eq.(\ref{eq: F(R_0)}) into Eq.(\ref{eq: action_desitter})
and setting $c_{1}=16$, we obtain the following action, 
\begin{equation}
S=\frac{2}{\kappa}\int d^{4}x\sqrt{-|g_{\mu\nu}|}R_{0}^{2}\label{eq: action_FR_4}
\end{equation}

\section{Conclusion}

In this work, we examined Born-Infeld-$f\left(R\right)$ gravity in
which the $f\left(R\right)$ function is considered in the determinantal
structure. We derived the equations of motions by introducing an auxiliary
metric without any restrictions in Eq.(\ref{eq: eom_g}) and Eq.(\ref{eq: eom_conn}). 

Then we analyzed this model for three cases. Firstly, we considered
the conformal assumption and we showed that the equations of motions
reduces to the result given in \citep{makarenko2014born}. According
to our results, we note that both actions given in Eq.(\ref{eq: action_f(R)})
and Eq.(\ref{eq: action_f(r)_2}) reduce to the same structure which
contains only the square of the Ricci scalar

Secondly, we use an exact $f\left(R\right)$ function that includes
positive and negative powers of the Ricci scalar with two constants.
We analyzed this assumption for three cases $F\left(R\right)=\gamma R+\frac{\epsilon}{R}$,
$F\left(R\right)=\gamma R$ and $F\left(R\right)=\frac{\epsilon}{R}$
respectively. The combined case leads to an action that contains Einstein's
theory plus the positive and negative power of the scalar curvature
(see Eq.(\ref{eq: action_FR_1})). Then the second condition is reduced
to the action (\ref{eq: action_FR_2}) which has already been studied
in \citep{chen2016modified}. For the last selection, the action (\ref{eq: action_FR_3})
become a similar structure to (\ref{eq: action_FR_1}) but the simpler
in the constants.

Finally, we looked for the existence of a maximally symmetric vacuum
solution for the gravitational field equations, so we assumed that
our theory should satisfy the de Sitter condition Eq.(\ref{eq: cons_curvature}).
Then we found the explicit structure of the $F\left(R_{0}\right)$
function and using this function the action (\ref{eq: action_f(r)_2})
was reduced to Eq.(\ref{eq: action_FR_4}) which contains only the
square of the constant scalar curvature. Note that the Schwarzschild-de
Sitter black hole solution is also constant curvature solution. Such
solutions will be discussed elsewhere.
\begin{acknowledgments}
The author wish to thank Sergei D. Odintsov and Shin'ichi Nojiri for
useful discussions and comments regarding the results presented in
this work. The work of S.K. has been supported by the Scientific and
Technological Research Council of Turkey (TUB\.{I}TAK) under the grant
number 2219.
\end{acknowledgments}

\bibliographystyle{apsrev4-2}
\bibliography{BIFR_2_dS}

\begin{thebibliography}{32}%
\makeatletter
\providecommand \@ifxundefined [1]{%
 \@ifx{#1\undefined}
}%
\providecommand \@ifnum [1]{%
 \ifnum #1\expandafter \@firstoftwo
 \else \expandafter \@secondoftwo
 \fi
}%
\providecommand \@ifx [1]{%
 \ifx #1\expandafter \@firstoftwo
 \else \expandafter \@secondoftwo
 \fi
}%
\providecommand \natexlab [1]{#1}%
\providecommand \enquote  [1]{``#1''}%
\providecommand \bibnamefont  [1]{#1}%
\providecommand \bibfnamefont [1]{#1}%
\providecommand \citenamefont [1]{#1}%
\providecommand \href@noop [0]{\@secondoftwo}%
\providecommand \href [0]{\begingroup \@sanitize@url \@href}%
\providecommand \@href[1]{\@@startlink{#1}\@@href}%
\providecommand \@@href[1]{\endgroup#1\@@endlink}%
\providecommand \@sanitize@url [0]{\catcode `\\12\catcode `\$12\catcode
  `\&12\catcode `\#12\catcode `\^12\catcode `\_12\catcode `\%12\relax}%
\providecommand \@@startlink[1]{}%
\providecommand \@@endlink[0]{}%
\providecommand \url  [0]{\begingroup\@sanitize@url \@url }%
\providecommand \@url [1]{\endgroup\@href {#1}{\urlprefix }}%
\providecommand \urlprefix  [0]{URL }%
\providecommand \Eprint [0]{\href }%
\providecommand \doibase [0]{https://doi.org/}%
\providecommand \selectlanguage [0]{\@gobble}%
\providecommand \bibinfo  [0]{\@secondoftwo}%
\providecommand \bibfield  [0]{\@secondoftwo}%
\providecommand \translation [1]{[#1]}%
\providecommand \BibitemOpen [0]{}%
\providecommand \bibitemStop [0]{}%
\providecommand \bibitemNoStop [0]{.\EOS\space}%
\providecommand \EOS [0]{\spacefactor3000\relax}%
\providecommand \BibitemShut  [1]{\csname bibitem#1\endcsname}%
\let\auto@bib@innerbib\@empty
\bibitem [{\citenamefont {Born}\ and\ \citenamefont
  {Infeld}(1934)}]{born1934foundations}%
  \BibitemOpen
  \bibfield  {author} {\bibinfo {author} {\bibfnamefont {M.}~\bibnamefont
  {Born}}\ and\ \bibinfo {author} {\bibfnamefont {L.}~\bibnamefont {Infeld}},\
  }\href {https://doi.org/10.1098/rspa.1934.0059} {\bibfield  {journal}
  {\bibinfo  {journal} {Proc. Roy. Soc. Lond. A}\ }\textbf {\bibinfo {volume}
  {144}},\ \bibinfo {pages} {425} (\bibinfo {year} {1934})}\BibitemShut
  {NoStop}%
\bibitem [{\citenamefont {Deser}\ and\ \citenamefont
  {Gibbons}(1998)}]{deser1998born}%
  \BibitemOpen
  \bibfield  {author} {\bibinfo {author} {\bibfnamefont {S.}~\bibnamefont
  {Deser}}\ and\ \bibinfo {author} {\bibfnamefont {G.~W.}\ \bibnamefont
  {Gibbons}},\ }\href {https://doi.org/10.1088/0264-9381/15/5/001} {\bibfield
  {journal} {\bibinfo  {journal} {Class. Quant. Grav.}\ }\textbf {\bibinfo
  {volume} {15}},\ \bibinfo {pages} {L35} (\bibinfo {year} {1998})}\BibitemShut
  {NoStop}%
\bibitem [{\citenamefont {Vollick}(2004)}]{vollick2004palatini}%
  \BibitemOpen
  \bibfield  {author} {\bibinfo {author} {\bibfnamefont {D.~N.}\ \bibnamefont
  {Vollick}},\ }\href {https://doi.org/10.1103/PhysRevD.69.064030} {\bibfield
  {journal} {\bibinfo  {journal} {Phys. Rev. D}\ }\textbf {\bibinfo {volume}
  {69}},\ \bibinfo {pages} {064030} (\bibinfo {year} {2004})}\BibitemShut
  {NoStop}%
\bibitem [{\citenamefont {Vollick}(2005)}]{vollick2005born}%
  \BibitemOpen
  \bibfield  {author} {\bibinfo {author} {\bibfnamefont {D.~N.}\ \bibnamefont
  {Vollick}},\ }\href {https://doi.org/10.1103/PhysRevD.72.084026} {\bibfield
  {journal} {\bibinfo  {journal} {Phys. Rev. D}\ }\textbf {\bibinfo {volume}
  {72}},\ \bibinfo {pages} {084026} (\bibinfo {year} {2005})}\BibitemShut
  {NoStop}%
\bibitem [{\citenamefont {Ba\~nados}\ and\ \citenamefont
  {Ferreira}(2010)}]{Banados2010eddington}%
  \BibitemOpen
  \bibfield  {author} {\bibinfo {author} {\bibfnamefont {M.}~\bibnamefont
  {Ba\~nados}}\ and\ \bibinfo {author} {\bibfnamefont {P.~G.}\ \bibnamefont
  {Ferreira}},\ }\href {https://doi.org/10.1103/PhysRevLett.105.011101}
  {\bibfield  {journal} {\bibinfo  {journal} {Phys. Rev. Lett.}\ }\textbf
  {\bibinfo {volume} {105}},\ \bibinfo {pages} {011101} (\bibinfo {year}
  {2010})},\ \bibinfo {note} {[Erratum: Phys.Rev.Lett. 113, 119901
  (2014)]}\BibitemShut {NoStop}%
\bibitem [{\citenamefont {Beltran~Jimenez}\ \emph {et~al.}(2018)\citenamefont
  {Beltran~Jimenez}, \citenamefont {Heisenberg}, \citenamefont {Olmo},\ and\
  \citenamefont {Rubiera-Garcia}}]{jimenez2018born}%
  \BibitemOpen
  \bibfield  {author} {\bibinfo {author} {\bibfnamefont {J.}~\bibnamefont
  {Beltran~Jimenez}}, \bibinfo {author} {\bibfnamefont {L.}~\bibnamefont
  {Heisenberg}}, \bibinfo {author} {\bibfnamefont {G.~J.}\ \bibnamefont
  {Olmo}},\ and\ \bibinfo {author} {\bibfnamefont {D.}~\bibnamefont
  {Rubiera-Garcia}},\ }\href {https://doi.org/10.1016/j.physrep.2017.11.001}
  {\bibfield  {journal} {\bibinfo  {journal} {Phys. Rept.}\ }\textbf {\bibinfo
  {volume} {727}},\ \bibinfo {pages} {1} (\bibinfo {year} {2018})}\BibitemShut
  {NoStop}%
\bibitem [{\citenamefont {Makarenko}\ \emph
  {et~al.}(2014{\natexlab{a}})\citenamefont {Makarenko}, \citenamefont
  {Odintsov},\ and\ \citenamefont {Olmo}}]{makarenko2014born}%
  \BibitemOpen
  \bibfield  {author} {\bibinfo {author} {\bibfnamefont {A.~N.}\ \bibnamefont
  {Makarenko}}, \bibinfo {author} {\bibfnamefont {S.}~\bibnamefont
  {Odintsov}},\ and\ \bibinfo {author} {\bibfnamefont {G.~J.}\ \bibnamefont
  {Olmo}},\ }\href {https://doi.org/10.1103/PhysRevD.90.024066} {\bibfield
  {journal} {\bibinfo  {journal} {Phys. Rev. D}\ }\textbf {\bibinfo {volume}
  {90}},\ \bibinfo {pages} {024066} (\bibinfo {year}
  {2014}{\natexlab{a}})}\BibitemShut {NoStop}%
\bibitem [{\citenamefont {Odintsov}\ \emph {et~al.}(2014)\citenamefont
  {Odintsov}, \citenamefont {Olmo},\ and\ \citenamefont
  {Rubiera-Garcia}}]{odintsov2014born}%
  \BibitemOpen
  \bibfield  {author} {\bibinfo {author} {\bibfnamefont {S.~D.}\ \bibnamefont
  {Odintsov}}, \bibinfo {author} {\bibfnamefont {G.~J.}\ \bibnamefont {Olmo}},\
  and\ \bibinfo {author} {\bibfnamefont {D.}~\bibnamefont {Rubiera-Garcia}},\
  }\href {https://doi.org/10.1103/PhysRevD.90.044003} {\bibfield  {journal}
  {\bibinfo  {journal} {Phys. Rev. D}\ }\textbf {\bibinfo {volume} {90}},\
  \bibinfo {pages} {044003} (\bibinfo {year} {2014})}\BibitemShut {NoStop}%
\bibitem [{\citenamefont {Makarenko}\ \emph
  {et~al.}(2014{\natexlab{b}})\citenamefont {Makarenko}, \citenamefont
  {Odintsov},\ and\ \citenamefont {Olmo}}]{Makarenko:2014nca}%
  \BibitemOpen
  \bibfield  {author} {\bibinfo {author} {\bibfnamefont {A.~N.}\ \bibnamefont
  {Makarenko}}, \bibinfo {author} {\bibfnamefont {S.~D.}\ \bibnamefont
  {Odintsov}},\ and\ \bibinfo {author} {\bibfnamefont {G.~J.}\ \bibnamefont
  {Olmo}},\ }\href {https://doi.org/10.1016/j.physletb.2014.05.024} {\bibfield
  {journal} {\bibinfo  {journal} {Phys. Lett. B}\ }\textbf {\bibinfo {volume}
  {734}},\ \bibinfo {pages} {36} (\bibinfo {year}
  {2014}{\natexlab{b}})}\BibitemShut {NoStop}%
\bibitem [{\citenamefont {Makarenko}(2014)}]{makarenko2014unification}%
  \BibitemOpen
  \bibfield  {author} {\bibinfo {author} {\bibfnamefont {A.~N.}\ \bibnamefont
  {Makarenko}},\ }\href {https://doi.org/10.1007/s10509-014-1955-2} {\bibfield
  {journal} {\bibinfo  {journal} {Astrophys. Space Sci.}\ }\textbf {\bibinfo
  {volume} {352}},\ \bibinfo {pages} {921} (\bibinfo {year}
  {2014})}\BibitemShut {NoStop}%
\bibitem [{\citenamefont {Makarenko}\ \emph
  {et~al.}(2014{\natexlab{c}})\citenamefont {Makarenko}, \citenamefont
  {Odintsov}, \citenamefont {Olmo},\ and\ \citenamefont
  {Rubiera-Garcia}}]{Makarenko:2014cca}%
  \BibitemOpen
  \bibfield  {author} {\bibinfo {author} {\bibfnamefont {A.~N.}\ \bibnamefont
  {Makarenko}}, \bibinfo {author} {\bibfnamefont {S.~D.}\ \bibnamefont
  {Odintsov}}, \bibinfo {author} {\bibfnamefont {G.~J.}\ \bibnamefont {Olmo}},\
  and\ \bibinfo {author} {\bibfnamefont {D.}~\bibnamefont {Rubiera-Garcia}},\
  }\href@noop {} {\bibfield  {journal} {\bibinfo  {journal} {TSPU Bulletin}\
  }\textbf {\bibinfo {volume} {12}},\ \bibinfo {pages} {158} (\bibinfo {year}
  {2014}{\natexlab{c}})}\BibitemShut {NoStop}%
\bibitem [{\citenamefont {Elizalde}\ and\ \citenamefont
  {Makarenko}(2016)}]{Elizalde:2016vsd}%
  \BibitemOpen
  \bibfield  {author} {\bibinfo {author} {\bibfnamefont {E.}~\bibnamefont
  {Elizalde}}\ and\ \bibinfo {author} {\bibfnamefont {A.~N.}\ \bibnamefont
  {Makarenko}},\ }\href {https://doi.org/10.1142/S0217732316501492} {\bibfield
  {journal} {\bibinfo  {journal} {Mod. Phys. Lett. A}\ }\textbf {\bibinfo
  {volume} {31}},\ \bibinfo {pages} {1650149} (\bibinfo {year}
  {2016})}\BibitemShut {NoStop}%
\bibitem [{\citenamefont {Chen}\ \emph {et~al.}(2016)\citenamefont {Chen},
  \citenamefont {Bouhmadi-L{\'o}pez},\ and\ \citenamefont
  {Chen}}]{chen2016modified}%
  \BibitemOpen
  \bibfield  {author} {\bibinfo {author} {\bibfnamefont {C.-Y.}\ \bibnamefont
  {Chen}}, \bibinfo {author} {\bibfnamefont {M.}~\bibnamefont
  {Bouhmadi-L{\'o}pez}},\ and\ \bibinfo {author} {\bibfnamefont
  {P.}~\bibnamefont {Chen}},\ }\href
  {https://doi.org/10.1140/epjc/s10052-016-3879-1} {\bibfield  {journal}
  {\bibinfo  {journal} {Eur. Phys. J. C}\ }\textbf {\bibinfo {volume} {76}},\
  \bibinfo {pages} {1} (\bibinfo {year} {2016})}\BibitemShut {NoStop}%
\bibitem [{\citenamefont {Banik}\ \emph {et~al.}(2018)\citenamefont {Banik},
  \citenamefont {Banik},\ and\ \citenamefont {Bhuyan}}]{banik2018dynamical}%
  \BibitemOpen
  \bibfield  {author} {\bibinfo {author} {\bibfnamefont {D.~K.}\ \bibnamefont
  {Banik}}, \bibinfo {author} {\bibfnamefont {S.~K.}\ \bibnamefont {Banik}},\
  and\ \bibinfo {author} {\bibfnamefont {K.}~\bibnamefont {Bhuyan}},\ }\href
  {https://doi.org/10.1103/PhysRevD.97.124041} {\bibfield  {journal} {\bibinfo
  {journal} {Phys. Rev. D}\ }\textbf {\bibinfo {volume} {97}},\ \bibinfo
  {pages} {124041} (\bibinfo {year} {2018})}\BibitemShut {NoStop}%
\bibitem [{\citenamefont {Guth}(1981)}]{guth1981inflationary}%
  \BibitemOpen
  \bibfield  {author} {\bibinfo {author} {\bibfnamefont {A.~H.}\ \bibnamefont
  {Guth}},\ }\href {https://doi.org/10.1103/PhysRevD.23.347} {\bibfield
  {journal} {\bibinfo  {journal} {Phys. Rev. D}\ }\textbf {\bibinfo {volume}
  {23}},\ \bibinfo {pages} {347} (\bibinfo {year} {1981})}\BibitemShut
  {NoStop}%
\bibitem [{\citenamefont {Linde}(1982)}]{linde1982new}%
  \BibitemOpen
  \bibfield  {author} {\bibinfo {author} {\bibfnamefont {A.~D.}\ \bibnamefont
  {Linde}},\ }\href {https://doi.org/10.1016/0370-2693(82)91219-9} {\bibfield
  {journal} {\bibinfo  {journal} {Phys. Lett. B}\ }\textbf {\bibinfo {volume}
  {108}},\ \bibinfo {pages} {389} (\bibinfo {year} {1982})}\BibitemShut
  {NoStop}%
\bibitem [{\citenamefont {Albrecht}\ and\ \citenamefont
  {Steinhardt}(1982)}]{albrecht1982cosmology}%
  \BibitemOpen
  \bibfield  {author} {\bibinfo {author} {\bibfnamefont {A.}~\bibnamefont
  {Albrecht}}\ and\ \bibinfo {author} {\bibfnamefont {P.~J.}\ \bibnamefont
  {Steinhardt}},\ }\href {https://doi.org/10.1103/PhysRevLett.48.1220}
  {\bibfield  {journal} {\bibinfo  {journal} {Phys. Rev. Lett.}\ }\textbf
  {\bibinfo {volume} {48}},\ \bibinfo {pages} {1220} (\bibinfo {year}
  {1982})}\BibitemShut {NoStop}%
\bibitem [{\citenamefont {Linde}(1983)}]{linde1983chaotic}%
  \BibitemOpen
  \bibfield  {author} {\bibinfo {author} {\bibfnamefont {A.~D.}\ \bibnamefont
  {Linde}},\ }\href {https://doi.org/10.1016/0370-2693(83)90837-7} {\bibfield
  {journal} {\bibinfo  {journal} {Phys. Lett. B}\ }\textbf {\bibinfo {volume}
  {129}},\ \bibinfo {pages} {177} (\bibinfo {year} {1983})}\BibitemShut
  {NoStop}%
\bibitem [{\citenamefont {Linde}(1990)}]{linde1990particle}%
  \BibitemOpen
  \bibfield  {author} {\bibinfo {author} {\bibfnamefont {A.~D.}\ \bibnamefont
  {Linde}},\ }\href@noop {} {\emph {\bibinfo {title} {{Particle physics and
  inflationary cosmology}}}},\ Vol.~\bibinfo {volume} {5}\ (\bibinfo
  {publisher} {CRC press},\ \bibinfo {year} {1990})\BibitemShut {NoStop}%
\bibitem [{\citenamefont {Linde}(1991)}]{linde1991axions}%
  \BibitemOpen
  \bibfield  {author} {\bibinfo {author} {\bibfnamefont {A.~D.}\ \bibnamefont
  {Linde}},\ }\href {https://doi.org/10.1016/0370-2693(91)90130-I} {\bibfield
  {journal} {\bibinfo  {journal} {Phys. Lett. B}\ }\textbf {\bibinfo {volume}
  {259}},\ \bibinfo {pages} {38} (\bibinfo {year} {1991})}\BibitemShut
  {NoStop}%
\bibitem [{\citenamefont {Linde}(1994)}]{linde1994hybrid}%
  \BibitemOpen
  \bibfield  {author} {\bibinfo {author} {\bibfnamefont {A.~D.}\ \bibnamefont
  {Linde}},\ }\href {https://doi.org/10.1103/PhysRevD.49.748} {\bibfield
  {journal} {\bibinfo  {journal} {Phys. Rev. D}\ }\textbf {\bibinfo {volume}
  {49}},\ \bibinfo {pages} {748} (\bibinfo {year} {1994})}\BibitemShut
  {NoStop}%
\bibitem [{\citenamefont {Kallosh}\ \emph {et~al.}(2002)\citenamefont
  {Kallosh}, \citenamefont {Linde}, \citenamefont {Prokushkin},\ and\
  \citenamefont {Shmakova}}]{kallosh2002gauged}%
  \BibitemOpen
  \bibfield  {author} {\bibinfo {author} {\bibfnamefont {R.}~\bibnamefont
  {Kallosh}}, \bibinfo {author} {\bibfnamefont {A.~D.}\ \bibnamefont {Linde}},
  \bibinfo {author} {\bibfnamefont {S.}~\bibnamefont {Prokushkin}},\ and\
  \bibinfo {author} {\bibfnamefont {M.}~\bibnamefont {Shmakova}},\ }\href
  {https://doi.org/10.1103/PhysRevD.65.105016} {\bibfield  {journal} {\bibinfo
  {journal} {Phys. Rev. D}\ }\textbf {\bibinfo {volume} {65}},\ \bibinfo
  {pages} {105016} (\bibinfo {year} {2002})}\BibitemShut {NoStop}%
\bibitem [{\citenamefont {Olmo}(2011)}]{Olmo:2011uz}%
  \BibitemOpen
  \bibfield  {author} {\bibinfo {author} {\bibfnamefont {G.~J.}\ \bibnamefont
  {Olmo}},\ }\href {https://doi.org/10.1142/S0218271811018925} {\bibfield
  {journal} {\bibinfo  {journal} {Int. J. Mod. Phys. D}\ }\textbf {\bibinfo
  {volume} {20}},\ \bibinfo {pages} {413} (\bibinfo {year} {2011})}\BibitemShut
  {NoStop}%
\bibitem [{\citenamefont {Olmo}\ and\ \citenamefont
  {Olmo}(2012)}]{olmo2012open}%
  \BibitemOpen
  \bibinfo {editor} {\bibfnamefont {G.~J.}\ \bibnamefont {Olmo}}\ and\ \bibinfo
  {editor} {\bibfnamefont {G.~J.}\ \bibnamefont {Olmo}},\ eds.,\ \href
  {https://doi.org/10.5772/45746} {\emph {\bibinfo {title} {{Open Questions in
  Cosmology}}}}\ (\bibinfo  {publisher} {InTech},\ \bibinfo {year}
  {2012})\BibitemShut {NoStop}%
\bibitem [{\citenamefont {Comelli}\ and\ \citenamefont
  {Dolgov}(2005)}]{comelli2005determinant}%
  \BibitemOpen
  \bibfield  {author} {\bibinfo {author} {\bibfnamefont {D.}~\bibnamefont
  {Comelli}}\ and\ \bibinfo {author} {\bibfnamefont {A.}~\bibnamefont
  {Dolgov}},\ }\href {https://doi.org/10.1088/1126-6708/2004/11/062} {\bibfield
   {journal} {\bibinfo  {journal} {JHEP}\ }\textbf {\bibinfo {volume}
  {2004}}\bibinfo  {number} { (11)},\ \bibinfo {pages} {062}}\BibitemShut
  {NoStop}%
\bibitem [{\citenamefont {Fiorini}(2013)}]{Fiorini:2013kba}%
  \BibitemOpen
\bibfield  {number} {  }\bibfield  {author} {\bibinfo {author} {\bibfnamefont
  {F.}~\bibnamefont {Fiorini}},\ }\href
  {https://doi.org/10.1103/PhysRevLett.111.041104} {\bibfield  {journal}
  {\bibinfo  {journal} {Phys. Rev. Lett.}\ }\textbf {\bibinfo {volume} {111}},\
  \bibinfo {pages} {041104} (\bibinfo {year} {2013})}\BibitemShut {NoStop}%
\bibitem [{\citenamefont {Nojiri}\ and\ \citenamefont
  {Odintsov}(2003)}]{nojiri2003modified}%
  \BibitemOpen
  \bibfield  {author} {\bibinfo {author} {\bibfnamefont {S.}~\bibnamefont
  {Nojiri}}\ and\ \bibinfo {author} {\bibfnamefont {S.~D.}\ \bibnamefont
  {Odintsov}},\ }\href {https://doi.org/10.1103/PhysRevD.68.123512} {\bibfield
  {journal} {\bibinfo  {journal} {Phys. Rev. D}\ }\textbf {\bibinfo {volume}
  {68}},\ \bibinfo {pages} {123512} (\bibinfo {year} {2003})}\BibitemShut
  {NoStop}%
\bibitem [{\citenamefont {Chiba}(2003)}]{chiba20031}%
  \BibitemOpen
  \bibfield  {author} {\bibinfo {author} {\bibfnamefont {T.}~\bibnamefont
  {Chiba}},\ }\href {https://doi.org/10.1016/j.physletb.2003.09.033} {\bibfield
   {journal} {\bibinfo  {journal} {Phys. Lett. B}\ }\textbf {\bibinfo {volume}
  {575}},\ \bibinfo {pages} {1} (\bibinfo {year} {2003})}\BibitemShut {NoStop}%
\bibitem [{\citenamefont {Dolgov}\ and\ \citenamefont
  {Kawasaki}(2003)}]{dolgov2003can}%
  \BibitemOpen
  \bibfield  {author} {\bibinfo {author} {\bibfnamefont {A.~D.}\ \bibnamefont
  {Dolgov}}\ and\ \bibinfo {author} {\bibfnamefont {M.}~\bibnamefont
  {Kawasaki}},\ }\href {https://doi.org/10.1016/j.physletb.2003.08.039}
  {\bibfield  {journal} {\bibinfo  {journal} {Phys. Lett. B}\ }\textbf
  {\bibinfo {volume} {573}},\ \bibinfo {pages} {1} (\bibinfo {year}
  {2003})}\BibitemShut {NoStop}%
\bibitem [{\citenamefont {Fay}\ \emph {et~al.}(2007)\citenamefont {Fay},
  \citenamefont {Tavakol},\ and\ \citenamefont {Tsujikawa}}]{fay2007f}%
  \BibitemOpen
  \bibfield  {author} {\bibinfo {author} {\bibfnamefont {S.}~\bibnamefont
  {Fay}}, \bibinfo {author} {\bibfnamefont {R.}~\bibnamefont {Tavakol}},\ and\
  \bibinfo {author} {\bibfnamefont {S.}~\bibnamefont {Tsujikawa}},\ }\href
  {https://doi.org/10.1103/PhysRevD.75.063509} {\bibfield  {journal} {\bibinfo
  {journal} {Phys. Rev. D}\ }\textbf {\bibinfo {volume} {75}},\ \bibinfo
  {pages} {063509} (\bibinfo {year} {2007})}\BibitemShut {NoStop}%
\bibitem [{\citenamefont {Nojiri}\ and\ \citenamefont
  {Odintsov}(2007)}]{nojiri2007introduction}%
  \BibitemOpen
  \bibfield  {author} {\bibinfo {author} {\bibfnamefont {S.}~\bibnamefont
  {Nojiri}}\ and\ \bibinfo {author} {\bibfnamefont {S.~D.}\ \bibnamefont
  {Odintsov}},\ }\href {https://doi.org/10.1142/S0219887807001928} {\bibfield
  {journal} {\bibinfo  {journal} {Int. J. Geom. Methods Mod. Phys.}\ }\textbf
  {\bibinfo {volume} {4}},\ \bibinfo {pages} {115} (\bibinfo {year}
  {2007})}\BibitemShut {NoStop}%
\bibitem [{\citenamefont {Nojiri}\ and\ \citenamefont
  {Odintsov}(2011)}]{nojiri2011unified}%
  \BibitemOpen
  \bibfield  {author} {\bibinfo {author} {\bibfnamefont {S.}~\bibnamefont
  {Nojiri}}\ and\ \bibinfo {author} {\bibfnamefont {S.~D.}\ \bibnamefont
  {Odintsov}},\ }\href {https://doi.org/10.1016/j.physrep.2011.04.001}
  {\bibfield  {journal} {\bibinfo  {journal} {Phys. Rept.}\ }\textbf {\bibinfo
  {volume} {505}},\ \bibinfo {pages} {59} (\bibinfo {year} {2011})}\BibitemShut
  {NoStop}%
\end{thebibliography}%

\end{document}